\documentclass[aps,prl,twocolumn,superscriptaddress,showpacs,showkeys,floatfix]{revtex4-1}
\usepackage{graphicx}
\usepackage{charter}
\usepackage{amssymb}
\usepackage{setspace}
\usepackage{amsmath}
\usepackage{csquotes}
\usepackage{float}
\usepackage {braket}
\usepackage{hyperref}
\usepackage{xcolor}
\usepackage{xfrac}
\definecolor{dark-red}{rgb}{0.4,0.15,0.15}
\definecolor{dark-blue}{rgb}{0.15,0.15,0.4}
\definecolor{medium-blue}{rgb}{0,0,0.5}
\hypersetup{
    colorlinks, linkcolor={dark-red},
    citecolor={dark-blue}, urlcolor={medium-blue}
}

\begin{document}


\title{An anisotropic local modification of crystal field levels in
  Pr-based pyrochlores: a muon-induced effect modelled using density
  functional theory}

\author{F.~R.~Foronda}
\affiliation{Oxford University Department of Physics, Clarendon Laboratory, Parks Road, Oxford, OX1 3PU, United Kingdom}

\author{F.~Lang}
\affiliation{Oxford University Department of Physics, Clarendon Laboratory, Parks Road, Oxford, OX1 3PU, United Kingdom}

\author{J.~S.~M\"{o}ller}
\altaffiliation[Current address: ]{
Laboratory for Solid State Physics, 
ETH Z\"urich, Z\"urich, Switzerland}
\affiliation{Oxford University Department of Physics, Clarendon
  Laboratory, Parks Road, Oxford, OX1 3PU, United Kingdom}

\author{T.~Lancaster}
\affiliation{Durham University, Centre for Materials Physics, South Road, Durham, DH1 3LE, United Kingdom}

\author{A.~T.~Boothroyd}
\affiliation{Oxford University Department of Physics, Clarendon Laboratory, Parks Road, Oxford, OX1 3PU, United Kingdom}

\author{F.~L.~Pratt}
\affiliation{ISIS Facility, Rutherford Appleton Laboratory, Chilton, Oxfordshire OX11 0QX, United Kingdom}

\author{S.~R.~Giblin}
\affiliation{School of Physics and Astronomy, Cardiff University,
Cardiff,
CF24 3AA, United Kingdom}

\author{D.~Prabhakaran}
\affiliation{Oxford University Department of Physics, Clarendon Laboratory, Parks Road, Oxford, OX1 3PU, United Kingdom}

\author{S.~J.~Blundell}
\email{s.blundell@physics.ox.ac.uk}
\affiliation{Oxford University Department of Physics, Clarendon Laboratory, Parks Road, Oxford, OX1 3PU, United Kingdom}

\date{\today}

\begin{abstract}
  Muon spin relaxation measurements on some quantum spin ice candidate
  materials, the insulating pyrochlores
  Pr$_2$\textit{B}$_2$O$_7$ (\textit{B} = Sn, Zr, Hf), have been
  performed for temperatures in the range 0.05--280\,K. The results
  are 
  indicative of a static distribution of magnetic
  moments which appears to grow on cooling and whose size at low
  temperatures is significantly larger than
  that expected for Pr nuclear moments.  
Using density functional theory we show how this effect can be
explained
via a hyperfine enhancement
  arising from a splitting of the non-Kramers doublet ground states on 
  Pr ions close to the muon which itself causes a highly anisotropic
  distortion field.  
We provide a quantitative
  relationship between this effect and the measured temperature
  dependence of the muon relaxation and discuss the relevance of these
  observations to muon experiments in other frustrated magnetic
  materials.
\end{abstract}

\pacs{76.75.+i, 75.10.$-$b, 75.40.Cx, 75.10.Jm}

\maketitle

The muon-spin relaxation ($\mu$SR) technique has been widely used as a
probe of exotic magnetic behavior in frustrated systems \cite{CarrettaKeren}.  A
crucial question for these experiments is to what extent the
presence of an implanted muon perturbs its local environment to such a
degree that the measured response reflects the nature of the local
distortions more than the physical behavior of the system under study.
To answer this question we have identified a worst-case scenario,
where the intrinsic magnetic behaviour is that of a quantum spin ice
originating from the magnetic moments of Pr$^{3+}$ ions.  The ground
state of this non-Kramers ion in a high symmetry site is particularly
susceptible to modification by the implanted muon.  By using
density-functional theory (DFT) and crystal-field (CF) calculations, we
show in this Letter how the observed behavior results from a highly
anisotropic distortion field induced by the implanted muon.

In pyrochlore oxides A$_2$B$_2$O$_7$, in which the magnetic A ions
occupy a lattice of corner-sharing tetrahedra, a variety of ground
states can be realised, including spin glasses and spin ices
\cite{GardnerRevModPhys82}.  Spin-ice behavior has been widely studied
in Dy$_2$Ti$_2$O$_7$ and Ho$_2$Ti$_2$O$_7$ (i.e.\ with A$=$Dy or Ho
and B$=$Ti)
and arises because the Ising spins are constrained by the CF
to point in or out of each tetrahedron and along the local
$\langle$111$\rangle$ axes
\cite{BramwellSci294-2001, CastelnovoNat451-2008}.  It has been
suggested that a new type of {\sl quantum} spin ice
\cite{GingrasMcClarty2014}
may be realised in
which A is a
lanthanide with fewer f electrons and a smaller magnetic moment, such
as Pr$^{3+}$ \cite{OnodaPRL105-2010, ZhouPRL101-2008}. This leads to 
a spatially extended 4f wave function with a greater
overlap with the oxygen 2p orbitals, as well as a weaker magnetic
dipolar interaction (proportional to the square of the moment size)
between nearest-neighbour sites. This can allow quantum
tunneling between different ice configurations, thereby converting
the material from a spin ice to a quantum spin liquid. Like its
classical counterpart, the most notable feature of this new state is that
it is predicted to host
unconventional excitations.   These are
linearly dispersive magnetic excitations (magnetic photons) which
offer the possibility of constructing a real lattice analogue of
quantum electromagnetism \cite{BentonPRB86-2012, LeePRBB86-2012}.
The compound Pr$_2$Ir$_2$O$_7$ has been identified as a highly
correlated
metallic spin liquid \cite{NakatsujiPRL96, TokiwaNatMat13} and a previous $\mu$SR investigation found
behavior that was interpreted as being induced by the muon
\cite{MacLaughlinPhysB404, MacLaughlinJPCS225}, although the extent of
the role of screening by the conduction electrons was not clear.
We now demonstrate that this effect can be also found in the
insulating
compounds  Pr$_2$B$_2$O$_7$ (\textit{B} = Sn, Zr, and Hf) which are
also candidate quantum spin ice systems, and we propose a mechanism
for the observed effect.

\begin{figure}[h]
\centering
\includegraphics[width=5cm]{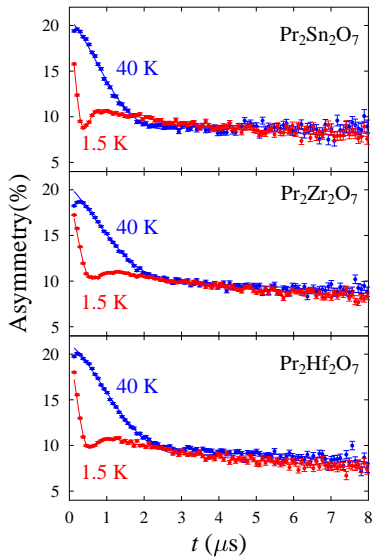} 
\caption{Zero-field $\mu$SR spectra at 1.5 K and 40 K. Fits are to a Kubo-Toyabe relaxation function.}
\label{ZF_RawData}
\end{figure}

Zero-field $\mu$SR measurements were carried out on polycrystalline
samples using the EMU spectrometer at the ISIS muon facility,
RAL. Data were taken in the temperature range  0.05--280\,K using a
$^3$He cryostat and $^3$He--$^4$He dilution refrigerator. The samples
were synthesised by standard solid-state reactions and confirmed by
x-ray diffraction to be single phase.
Representative raw spectra from $\mu$SR measurements taken in zero
applied field at 1.5\,K and 40\,K are shown in Fig.~\ref{ZF_RawData}. 
At 1.5\,K 
all compounds show a Kubo-Toyabe relaxation function (an initially Gaussian
depolarization which recovers to a $\frac{1}{3}$ constant long-time
tail) which can be
understood as resulting from a distribution of randomly
oriented, static magnetic moments
\cite{HayanoPRB20}.  
The static moments lead to a Gaussian distribution of magnetic fields
at the muon site of rms width $B_{\rm rms}=\Delta/\gamma_{\mu}$
where $\gamma_{\mu}=2 \pi \times 135.5 \text{MHz T}^{-1}$ is the muon
gyromagnetic ratio.  However, the value of $\Delta$ extracted at low
temperature is too large to originate simply from nuclear spins.
We fit our data  to a product of a Gaussian Kubo-Toyabe function $\text
G_{KT}(\Delta,t)$ and a weakly relaxing exponential $e^{-\lambda t}$,
the latter component to take into account some slow dynamics of 
the magnetic moments, and this fit function can be used across the entire
temperature range studied.
All samples show a small, rapidly-relaxing fraction which we interpret as a
muonium state  (and is responsible for the
negative curvature in the asymmetry data at very short times), but we
focus on the majority fraction in the subsequent discussion.

 \begin{table}[H]
 \caption{Lattice constants $a$ (at 300\,K), 
$B_{\rm rms}$ and fitted energy gaps for  Pr$_2$\textit{B}$_2$O$_7$ (\textit{B} = Zr, Hf,
Sn and Ir). Pr$_2$Ir$_2$O$_7$ data taken from \cite{MacLaughlinPhysB404, NakatsujiPRL96}. \label{Pr-constants}} 
 \begin{ruledtabular}
    \begin{tabular}{l c c c c}
    compound               & Zr & Hf & Sn & Ir \\
\hline
    $a$ (\AA)              &  $10.7386(2)$  & $10.7177(2)$  & $10.6055(2)$  &$10.3940(4)$  \\
    $B_{\rm rms}$ (mT) & $3.76(6)$  & $4(2)$  & $5.65(5)$ & $8.9(4)$    \\
$\epsilon_1$ (meV) & $3.1(5)$ & $2(2)$ & $6(1)$ & $0.5(1)$ \\
$\epsilon_2$ (meV) & $0.6(1)$ & $0.5(7)$ & $0.6(1)$ & $0.1(1)$ \\
    \end{tabular}
 \end{ruledtabular}
 \end{table}

\begin{figure}[h]
\centering
\includegraphics*[trim = 0cm 0cm 0cm 0cm, clip=true, width=7.0cm]{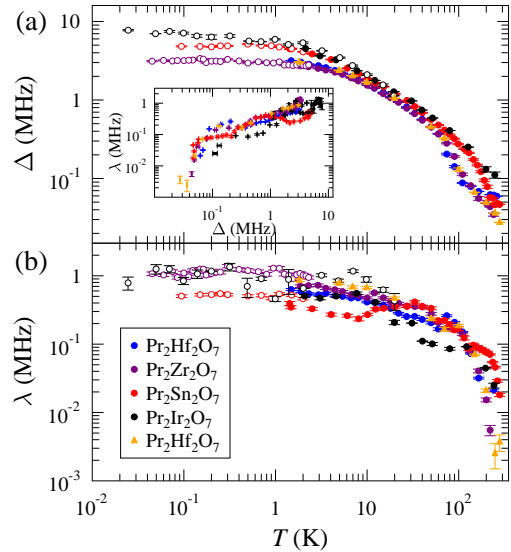} 
\caption{Temperature dependences of (a) $\Delta$
  and (b) $\lambda$ in Pr-based pyrochlores
  (helium cryostat, filled circles; dilution fridge, open
  circles). The two data sets for Pr$_2$Hf$_2$O$_7$ were taken with
  the same sample but in separate experiments to check reproducibility. The Pr$_2$Ir$_2$O$_7$ data are taken from \cite{MacLaughlinPhysB404}.}
\label{ZF_fits}
\end{figure}

The
temperature dependences of the static width $\Delta$ and dynamic
relaxation rate $\lambda$ are shown in Fig.~\ref{ZF_fits}. No
magnetic transition can be seen throughout the measured temperature
range; both parameters evolve smoothly, increasing steadily as
the samples are cooled and the increase in $\Delta$ can be interpreted
as magnetic moments which grow with decreasing temperature. 
The two parameters $\lambda$ and
$\Delta$ roughly track each other (see inset to Fig.~\ref{ZF_fits}), suggesting that the
dynamics are related to these growing moments. 
The values of  $B_{\rm rms}$ extrapolated to zero
temperature (see Table \ref{Pr-constants}) are found to be an order
of magnitude larger than expected from $^{141}$Pr nuclear moments
\cite{MacLaughlinPhysB404, MacLaughlinJPCS225} and are larger for
smaller lattice constants $a$. Moreover, at low
temperature we find that the relaxation decouples much more quicky in
applied longitudinal fields than would be expected from the fitted
values of $B_{\rm rms}$.

In these quantum spin ice system we do not expect any static
electronic moments at any temperature.  Nuclear moments would be
expected to give a small temperature-independent $\Delta$, but the
observed behavior in Fig.~\ref{ZF_fits}(a) is both strongly
temperature-dependent and at low temperatures very large.  A likely
explanation comes via a hyperfine enhancement of the Pr nuclear
moments (as proposed for Pr$_2$Ir$_2$O$_7$ 
\cite{MacLaughlinPhysB404, MacLaughlinJPCS225} and discussed in more detail below), 
but this mechanism requires a non-magnetic
(singlet) ground state.  In the pyrochlore structure the Pr$^{3+}$
(4f$^2$) ground state is a well-isolated non-Kramers doublet
(confirmed in Pr$_2$Sn$_2$O$_7$ and Pr$_2$Zr$_2$O$_7$ by neutron spectroscopy
\cite{PrincepPRB88,KimuraNC}) and this could be split by the distortion
introduced by the muon.
A muon-induced perturbation of the CF has been
suggested previously in Pr-based intermetallics 
\cite{FeyerhermZPB99, TashamaPRB56}, similarly associated with the splitting of the non-Kramers doublet. 
Since our data in insulating pyrochlores look very similar to that in metallic
Pr$_2$Ir$_2$O$_7$ we conclude that this mechanism is not susceptible
to screening effects.  In fact the carrier density in
Pr$_2$Ir$_2$O$_7$
is found to be rather low (estimated to be $2.6\times10^{20}$ cm$^{-3}$, i.e.\ $\approx
0.02$ conduction electrons per Pr,
from Hall effect measurements \cite{NakatsujiPRL96}).  Moreover, 
Pr$_2$Ir$_2$O$_7$ is believed to have a Fermi node
at the $\Gamma$ point \cite{2014arXiv1403.5255S}, so that conduction electrons can only screen
effectively at very long wavelengths. Both considerations
allow us to rationalise the insensitivity of this effect to the degree of
metallicity.

\begin{figure}[h]
\centering
\includegraphics*[width=7cm]{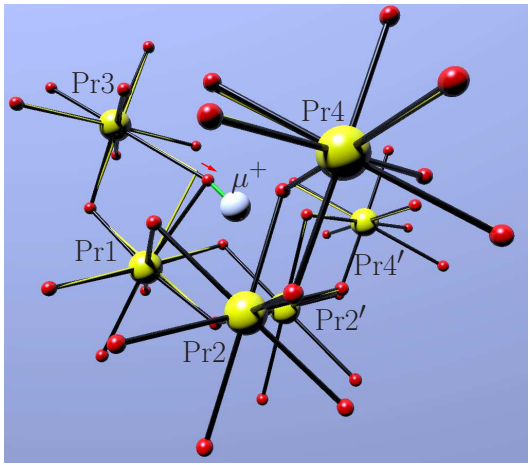} 
\caption{Muon stopping site (white sphere) and atomic positions in
  Pr$_2$Sn$_2$O$_7$ as calculated by DFT. Oxygen sites shown in red;
  Pr ions in yellow and bonds are shown as solid black lines. Bonds of
  the unperturbed lattice (yellow lines) are included for
  comparison. Pr ions are numbered in order of separation from $\mu$
  with Pr1 being the closest.  The pair Pr2 and Pr2' are equidistant from the muon, as are
  Pr4 and Pr4'.}
\label{DFT}
\end{figure}

Although there has been previous evidence for a muon-induced effect in Pr-containing systems with a
non-Kramers doublet \cite{FeyerhermZPB99, TashamaPRB56, MacLaughlinPhysB404}, the nature of the effect has not been explored in
detail.
To address this issue we have used DFT calculations to determine the
muon location and assess the effect of the muon on the
local crystal structure and the CF of nearby
Pr ions.
 The DFT calculations reported here were conducted with the plane-wave
 \textit{Quantum Espresso}~\cite{Giannozzi} program and utilized the generalized
 gradient approximation (GGA) exchange-correlation functional of
 Perdew, Burke and Ernzerhof \cite{PerdewPRL77} to locate potential
 muon sites. Ions were modeled using ultrasoft pseudopotentials and
 the muon was modeled by a norm-conserving hydrogen
 pseudopotential.  This technique is known to give 
reliable results for muon sites in condensed matter systems
\cite{Moller2013, Bernardini2013, f4bimnn2013}.
The effect of including the 4f electrons in valence in the Pr pseudopotential (which is computationally challenging) on the determined muon sites and bulk lattice parameters was found to be negligible and hence a Pr pseudopotential with the 4f electrons in the core was employed for the calculations described below.
The calculations were performed for
 Pr$_2$Sn$_2$O$_7$ in a supercell consisting of a single
 conventional unit cell containing 88 atoms, and with the total energy
 converged to at least $1 \times 10^{-6}$ Ry/atom (where Ry is the
 Rydberg constant). A convergence test yielded the suitable
 wavefunction and charge density cutoffs of 50\,Ry and 300\,Ry
 respectively on a $2\times2\times2$ Monkhorst-Pack
 $k$-space grid, which were then used in all
 subsequent calculations. The calculated atomic positions and lattice
 parameter of the unperturbed bulk were within 2\% of the experimental
 values reported in \cite{KennedyJSSC130}, demonstrating an excellent
 agreement with x-ray and neutron powder experiments. To determine the
 stopping site, a muon was introduced on a grid of low-symmetry
 positions and the system was allowed to relax until all forces were
 below $10^{-3}$ Ry/a.u. and the change in energy between iterations
 was less than $10^{-4}$ Ry. The calculations presented here focus
 solely on the diamagnetic muon state, for which the unit cell has a
 total charge of +1. The final relaxed positions were found to be the
 same in spin-polarized and non-polarized calculations.

Three potential stopping sites were identified. However, as two of
these required configurations that were $\sim0.45$\,eV and 0.9\,eV
higher in energy than that of the lowest state we conclude that the
latter is the most plausible stopping site in our real system. In this
scenario the muon forms a O--H type bond of length $\approx$1\,\AA.  It
is bonded to an O$^{2-}$ ion which also bonds to two Pr
ions, labelled Pr1 and Pr3 in Fig.~\ref{DFT} (the Pr ions
in Fig.~\ref{DFT} are numbered in order of separation from the muon,
with Pr1 being the closest). The implanted muon 
results in an anisotropic distortion of the crystal lattice.  The muon
pulls an O$^{2-}$ away from the ion Pr3, resulting in a greatly
extended Pr3--O bond.  The Pr1--O bond is only slightly changed in
length, but is bent round, resulting in an anisotropic distribution of
O$^{2-}$ ions around Pr1.   As shown below, the largest change in
CF ground state is found for Pr3, but we note that the
environment around the ions labelled Pr2, Pr2', Pr4 and Pr4' are 
more gently modified.

\begin{table}[ht]
 \caption{Parameters derived from DFT
   calculations of muon-induced lattice perturbations in
   Pr$_2$Sn$_2$O$_7$. Values are shown for the four nearest-neighbour
   Pr ions.} 
 \begin{ruledtabular}
    \begin{tabular}{l c c c c}
    Pr atom               & 1 & 2,2' & 3 & 4,4' \\
\hline
    Pr--$\mu$ separation (\AA)               & $2.7$    & $3.2$ &
    $4.1$  & $4.7$  \\
    Relative contribution & 1.0 & 0.62 & 0.30 & 0.19 \\
    Distortion of PrO$_8$ unit (\AA)  & $0.23$   & $0.07$ & $0.56$ & $0.09$  \\
    $\epsilon$ (meV)              & $4.8$  & $1.3$ &  $11.4$   & $4.0$\\
    \end{tabular}
 \end{ruledtabular}
 \label{DFTresults}
 \end{table}

\begin{figure}[ht]
\centering
\includegraphics*[trim = 0cm 0cm 0cm 0cm, clip=true, width=6.5cm]{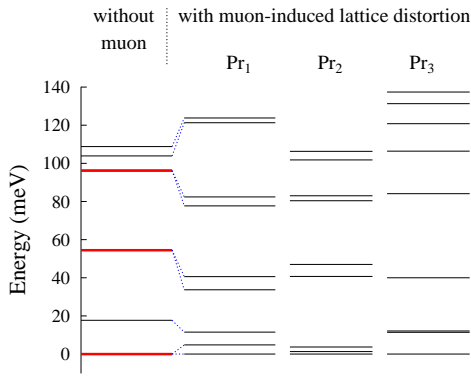} 
\caption{Calculated CF levels of Pr$^{3+}$ with and without
  muon-induced lattice distortion in Pr$_2$Sn$_2$O$_7$. Energies are
  calculated for three  Pr$^{3+}$ sites with varying degrees of distortion (see table \ref{DFTresults}) as determined by DFT calculations. Bold red lines indicate doublets; plain black lines are singlets.}
\label{CF_levels}
\end{figure}

We quantify the relative distortion of a PrO$_8$
unit  as the rms of the differences in Pr--O bond lengths
between the perturbed and unperturbed lattice and these are listed in
Table~\ref{DFTresults} for each of the nearest sites. 
We have calculated the CF levels for all nearby
PrO$_8$ environments, taking into account the spatial arrangement of
the eight nearest-neighbour oxygen anions around each Pr (the electrostatic
field due to the muon itself was also included initially, but it was
found to
make little difference, and so was neglected for the
calculations described here).  
We used a point-charge model, with effective charges on the O1 and O2
sites chosen to reproduce the measured CF spectrum \cite{PrincepPRB88}. For each Pr
site, the presence of the muon 
splits
the non-Kramers ground state doublet into two singlets
(Fig.~\ref{CF_levels}).
As expected, the largest splitting ($\epsilon$) of the singlets
is found for Pr3 (the splittings are listed in Table \ref{DFTresults}).
These calculations show that the most perturbed Pr ion is not the
nearest to the muon (there are three closer Pr ions that are
significantly less perturbed), reflecting the highly anisotropic
nature of the induced distortion field.
Thus we conclude from these calculations 
that the muon is surrounded by a number
of close Pr ions in which the CF splitting varies considerably.

\begin{figure}[h]
\centering
\includegraphics*[trim = 0cm 0cm 0cm 0cm, clip=true, width=6.5cm]{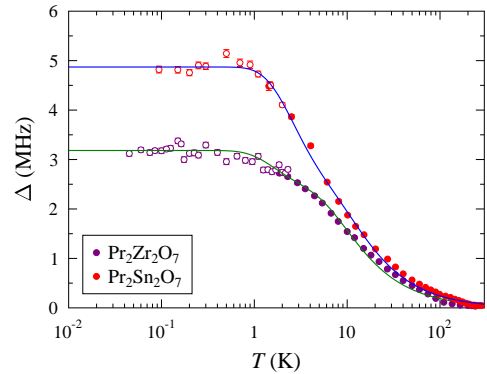} 
\caption{Static relaxation rate $\Delta$ in Pr$_2$Sn$_2$O$_7$ and
  Pr$_2$Zr$_2$O$_7$ fitted with  the model described in the text.}
\label{Bleaney_fit}
\end{figure}

We now turn to the hyperfine enhancement of the Pr nuclear
spins caused by these CFs.  One can consider a
two-state model due to Bleaney \cite{BleaneyPhysica69} 
in which the non-Kramers doublet is split into two singlets $\Ket{G}$ and $\Ket{E}$ by a small energy $\epsilon$. For a nucleus with spin $\boldsymbol I$ the hamiltonian takes the form
\begin{equation}\label{hamiltonian}
\mathcal{H}=\mathcal{H_{\mathrm X}}+g_J \mu_{\mathrm B}\boldsymbol B \cdot \boldsymbol J+A_J\boldsymbol J \cdot \boldsymbol I-g_I\mu_{\mathrm B}\boldsymbol B \cdot \boldsymbol I.
\end{equation}
Here the field $\boldsymbol B$ is applied along the z--direction and
$\mathcal{H_{\mathrm x}}$ accounts for the CF and the
splitting $\epsilon$. There is an electronic matrix element
${\alpha=\Bra{E}\hat{J}_{\text z}\ket{G}}$ where $\hat{J}_z$ is the
electronic angular momentum. This model allows an estimate of the
magnetic moment $m=k_{\rm B}T \left( \partial  \ln Z /\partial B \right)_T$ where $Z$ is the partition function and yields
\begin{equation}\label{moment}
m=g_I\mu_{\mathrm B}I_z+g_J\mu_{\mathrm B}\alpha \sin{\theta} \tanh \left( \frac{\epsilon}{2 \cos \theta k_{\mathrm B}T}\right)
\end{equation}
where $\tan \theta=2 \alpha (g_I\mu_{\mathrm B}B_{\mathrm
  z}+A_JI_z)/\epsilon$. In zero-field $\mu$SR we
take $B_z=0$ and hence 
\begin{equation}\label{ZFmoment}
m=m_0+\frac{\eta}{\tilde{\epsilon}} \tanh \left( \frac{\tilde{\epsilon}}{k_{\mathrm B}T}\right)
%
%
\end{equation}
where $m_0=g_I\mu_{\mathrm B}I_z$,
$\eta=g_J\mu_{\mathrm B}\alpha^2 A_JI_{\mathrm
    z}$ and
$\tilde{\epsilon}=\sqrt{(\epsilon/2)^2+(\alpha A_J
    I_z)^2}$ (where $\alpha^2\leq J^2$).  Taking $\Delta \propto m$ with $A_{\mathrm
  J}/h=1.093$\,GHz \cite{BleaneyPhysica69}  and $I_z=\frac{5}{2}$ allows an estimate of the upper
bound of $\epsilon$. 
The muon is coupled to many neighbouring moments by the
dipole-dipole interaction which is proportional to $r^{-3}$ (the
relative contribution of this coupling for each site is listed in
Table~\ref{Pr-constants}, assuming a $r^{-3}$ dependence).  
For simplicity, we choose a model in which
there are dominant contributions to $\Delta$ 
from two nearby moments which act in quadrature.  We note that
Eq.~\ref{ZFmoment} implies that (neglecting the $m_0$ component and
for $\epsilon\gg \alpha^2A_JI_z$) the
enhanced moment is approximately inversely proportional to
$\epsilon$ at low temperature, 
and therefore we expect the response to be dominated by
nearby sites with small splittings.  
The zero-field data sets for both Sn and Zr compounds are
found to fit well to this two-component model, 
see Fig.~\ref{Bleaney_fit}.  
The fitted values for all compounds are listed in Table~\ref{Pr-constants}. 
We note that these values are within the same order of magnitude as our
estimated splittings for the nearest neighbour sites Pr1 and Pr2/Pr2'
for Pr$_2$Sn$_2$O$_7$. Given the sensitivity of the calculations to
the precise distortion field, the restriction to two components,
together with the limitations of the point-charge
model of the CF, we believe this agreement is well within the
inherent uncertainties.

To summarize, we have performed $\mu$SR measurements on Pr-based
pyrochlores Pr$_2$\textit{B}$_2$O$_7$ (\textit{B} = Sn, Zr, Hf) in
which any quantum spin-ice type behavior is masked
by the presence of static, hyperfine-enhanced nuclear moments
with weak dynamics which dominate the muon relaxation.  
This effect is due to a
distribution of splittings of the non-Kramers doublet ground
states of nearby Pr ions resulting from a highly anisotropic
distortion field induced by the implanted muon.  These observations
show that, in certain circumstances, $\mu$SR experiments can measure a
response which is dominated by the local distortion resulting from the
implanted probe.  This particular case is, however, very unusual
since it relies on a splitting of a non-Kramers doublet and would be
inoperable in systems in which a ground state degeneracy was
protected from such perturbations, such as Dy$_2$Ti$_2$O$_7$.  
Nevertheless, we expect that in muon experiments on other
pyrochlore oxides a very similar anisotropic distortion field will
inevitably be
present, even if its effect is much more benign.

This work is supported by EPSRC (UK).  The computations were performed on the Iridis cluster operated by the E-Infrastructure South Initiative.
We thank Davide Ceresoli for helpful discussions and the provision of
a Pr pseudopotential with 4f electrons in valence for cross-checking
our results, and S.~P.~Cottrell at ISIS for technical assistance with the experiments.

\end{document}